\documentclass[twocolumn,amsmath,aps,fleqn]{revtex4}
\usepackage{amssymb}
\usepackage{amsmath}
\usepackage[usenames]{color}
\usepackage{graphicx}

\usepackage{hyperref}

 \providecommand{\eprint}[1]{\href{http://arxiv.org/abs/#1}{#1}}
 \providecommand{\adsurl}[1]{\href{#1}{ADS}}
 
\providecommand{\url}[1]{\href{#1}{#1}}

\newcommand{\himpc}{\ensuremath{\,h\,{\rm Mpc}^{-1}}}
\newcommand{\ud}{\mathrm{d}}
\providecommand{\eprint}[1]{\href{http://arxiv.org/abs/#1}{#1}}
 \providecommand{\adsurl}[1]{\href{#1}{ADS}}
 
\providecommand{\url}[1]{\href{#1}{#1}}

\newcommand\eq[1]{Eq.~(\ref{#1})}
\newcommand\eqs[2]{Eqs.~(\ref{#1}) and (\ref{#2})}

\newcommand{\beq}{\begin{equation}}
\newcommand{\eeq}{\end{equation}}
\newcommand{\bea}{\begin{eqnarray}}
\newcommand{\eea}{\end{eqnarray}}

\newcommand{\CDM}{\rm CDM}

\newcommand{\mnras}{Mon. Not. R. Astron. Soc.}
\newcommand{\physrep}{Phys.~Rep.}
\newcommand{\apjs}{Astrophys. J. S.}

\newcommand{\lsim}{\,\raise 0.4ex\hbox{$<$}\kern -0.8em\lower 0.62ex\hbox{$\sim$}\,}

\def\Cnoise{C_0^{\rm noise}}
\def\Clnoise{C_\ell^{\rm noise}}

\def\fsky{f_{\rm sky}}
\def\fcover{f^{\rm cover}}

\def\l{\ell}

\def\Cnoise{C_0^{\rm noise}}
\def\Clnoise{C_\l^{\rm noise}}

\def\I{{\bf I}}

\def\P{{\cal P}}

\def\Tsys{T_{\rm sys}}

\def\Omegamap{\Omega_{\rm map}}

\def\I{{\cal I}}

\begin{document}
\title{Forecasted 21 cm  constraints on compensated isocurvature perturbations}
\author{Christopher Gordon}
\affiliation{Beecroft Institute of Particle Astrophysics and Cosmology, University of Oxford, UK}
\author{Jonathan R.~Pritchard}\thanks{Hubble Fellow}
\affiliation{Institute for Theory \& Computation, Harvard-Smithsonian Center for Astrophysics, USA}
	

\label{firstpage}

\begin{abstract}
A ``compensated'' isocurvature perturbation 
consists of an overdensity (or underdensity) in the cold dark matter which is completely cancelled out by a corresponding underdensity (or overdensity) in the baryons.
 Such a configuration may be generated by a curvaton model of inflation if the cold dark matter is created before curvaton decay and the baryon number is created by the curvaton decay (or vice-versa). Compensated isocurvature perturbations, at the level producible by the curvaton model, have no observable effect on  cosmic microwave background anisotropies or on galaxy surveys. They can be detected through their effect on the distribution of neutral hydrogen between redshifts 30 to 300 using 21 cm absorption observations. However, to obtain a good signal to noise ratio, very large 
observing arrays are needed. We estimate that a fast Fourier transform telescope would need a total collecting  area of about 20 square kilometers to detect a curvaton generated compensated isocurvature perturbation at more than 5 sigma significance. 
\end{abstract}

\maketitle

\section{Introduction}
Current observations can be fit by a simple 6 parameter model of a flat $\Lambda$CDM Universe with a nearly scale invariant, adiabatic, and Gaussian primordial perturbation spectrum (see for example \cite{spergel07,komatsu08}).
The primordial perturbations may have been generated by a period of accelerated expansion in the early Universe, known as {\em inflation\/} (see for example \cite{lidlyt00}). Inflation can be driven by a potential dominated scalar field (the {\em inflaton}). The  current Universe contains cold dark matter ($\CDM$), baryons ($b$), neutrinos ($\nu$) and photons ($\gamma$). They can arise from the decay products of the inflaton. 
We assume the recent accelerated expansion of the Universe is been driven by a cosmological constant which is unperturbed.
Density perturbations in the other components can be inherited from the density perturbations in the inflaton which arise from vacuum fluctuations that are amplified to scales larger than the Hubble horizon during inflation. The perturbations are known as {\em adiabatic\/} if 
\beq
\left( \frac{ \delta\rho_i}{\dot{\rho}_i} - \frac{ \delta\rho_j}{\dot{\rho}_j}\right)=0
\eeq
 where $\delta \rho$ is the perturbation in the density, a dot indicates a derivative with respect to time and  $i$ and $j$ are each one of ($\CDM$, $b$, $\nu$, and $\gamma$). Non-adiabatic (also known as {\em isocurvature\/} or {\em entropy\/}) perturbations cannot arise if all the constituents of the Universe are the result of the decay of a single inflaton (see for example \cite{wands00,weinberg04}). In order to generate isocurvature perturbations there has to be more than one light degree of freedom present during inflation. In the {\em curvaton\/} model \cite{mortak01,lytwan02} , the perturbations generated  by the inflaton are negligibly small. But, there is a second light field (known as the {\em curvaton\/}) which also acquires perturbations, from vacuum fluctuations, during  inflation. At the end of inflation, the inflaton decays into radiation whose density decreases like $a^{-4}$, where $a=1/(1+z)$ is the scale factor and $z$ is the redshift. After inflation, when the Hubble parameter drops below the curvaton mass, the curvaton oscillates in the well of its potential  and its density decreases like $a^{-3}$
 and so its energy density relative to the radiation increases as $a$. When the Hubble parameter drops below the curvaton decay rate, the curvaton decays and its decay products are responsible for the observed density perturbations. If all of the constituents of the current Universe originate from the curvaton, then the resulting perturbations will be adiabatic. However, if some of the constituents originate from before the curvaton has non-negligible energy density and some from the curvaton, or its decay products, then  isocurvature modes are possible \cite{lytungwan03}. The magnitude of the adiabatic mode can be given in terms of the curvature perturbation in the constant density gauge, $\zeta$ (see for example \cite{bardeen80,barstetur83,lidlyt00}). In the cases we consider it is virtually equivalent to the curvature perturbation (or its negative, depending on the sign convention) in the comoving gauge, $\cal R$.  The isocurvature term between constituent $i$ and $j$ can be quantified by 
\beq
\label{isodef}
S_{i,j} = -3 H  \left( \frac{ \delta\rho_i}{\dot{\rho}_i} - \frac{ \delta\rho_j}{\dot{\rho}_j}\right)
=\frac{ \delta\rho_i}{(1+w_i){\rho}_i} - \frac{ \delta\rho_j}{(1+w_j){\rho}_j}
\eeq
where $H$ is the Hubble parameter, the second equality follows from the continuity equation, and $w$ is zero for baryons and CDM and a third for photons and neutrinos. If the CDM 
is created just after inflation, before the curvaton has non-negligible energy density
(as in the case of Wimpzillas \cite{damvil96,lytrobsmi98,chukolrio98}), then \cite{lytungwan03}
\beq
S_{\CDM,\gamma}=-3\zeta\,.
\label{scdm}
\eeq
Such a large negatively correlated isocurvature perturbation on its own is ruled out by current data \cite{gorlew03}. However, it may be offset by a positively correlated isocurvature perturbation between the baryons and the photons.
If the curvaton decay generates the baryon number (e.g.\ baryon number from the decay of a right-handed sneutrino curvaton \cite{hammuryan02}),  then \cite{lytungwan03}
\beq
S_{b,\gamma} = 3\frac{1-r}{r}\zeta\,.
\label{sb}
\eeq
where $r\approx\rho_{\rm curvaton}/\rho_{\rm total}$ at the time of curvaton decay.
Observations of the anisotropy in the cosmic microwave background (CMB) are unable to distinguish between a $S_{b,\gamma} $ and a $S_{\CDM,\gamma} $ isocurvature perturbation \cite{gorlew03}. Also, as discussed in 
Sec.~\ref{effects}, they cannot be distinguished with galaxy redshift surveys either. However, the effect of a CDM isocurvature mode is $\Omega_{\CDM}/\Omega_b$ times larger than that of a baryon isocurvature mode with the same amplitude. Where $\Omega_i$ is the density of component $i$ divided by the critical density.
It follows that  there is a,  currently unconstrained, compensated mode
where
\bea
S_{\CDM,\gamma}&=&\I \zeta, \label{comp1} \\  S_{b,\gamma}&=& -\frac{\Omega_{\CDM}}{\Omega_b} \I \zeta\, . \label{comp}
\eea
 The case where the CDM is created before curvaton decay and the baryon number by curvaton decay corresponds to $\I=-3$.

As pointed out in \cite{barloe05,lewcha07}, high redshift 21 cm observations can distinguish between $S_{\CDM,\gamma}$ and $S_{b,\gamma}$. In this article we estimate how large an array would be needed to detect 
 a curvaton generated compensated isocurvature mode.
The layout of the article is as follows. In Sec.~\ref{effects}, \ref{effectsgal}, and \ref{effects21}. the effect of compensated isocurvature perturbations is discussed. In Sec.~\ref{forecasts} forecasts for how well the compensated isocurvature perturbations can be constrained are made for possible future 21 cm experiments. The conclusions and discussion are  given in Sec.~\ref{conclusions}.

\section{Effect of Compensated Isocurvature on the CMB}
\label{effects}
In \cite{gorlew03} it was shown that the compensated mode is not observable in the CMB anisotropy. Here we go through the argument in a bit more detail and highlight the effect of baryon pressure.
The temperature perturbation in the CMB is given by the sum of the temperature perturbation from the adiabatic mode and that from the compensated mode:
\beq
\delta T =\delta \left.T\right|_{\I=0}+\delta \left.T\right|_{\zeta =0}\,.
\eeq
The compensated isocurvature mode has all perturbations set to zero except the baryon and CDM densities which have $\delta \rho_{\CDM} = -\delta \rho_b$ and so the total density is unperturbed.
In the compensated mode, the divergence of the baryon fluid velocity is sourced by the baryon  density (see for example \cite{maber95}). But, the sourcing term is of the form $c_s^2 k^2 \delta\rho_b/\rho_b$ where the baryon sound speed is give by \cite{maber95} 
 \beq
 c_s^2 = \frac{k_{B}T_b}{\mu}\left(1-\frac{1}{3}\frac{\ud\ln T_b}{\ud\ln a }\right)
 \eeq
 and $T_b$ is the baryon temperature (which is equal to the photon temperature until $z\sim 300$) and $\mu$ is the mean molecular weight (including free electrons and all ions in H and He). Only wave numbers for which 
 $k^2c_s^2 > {\cal H}^2$ (where $\cal H$ is the comoving Hubble parameter) are affected by the baryon sound speed.
 At last scattering, this corresponds to $k>200 h$Mpc$^{-1}$ which corresponds to angular scales of the observed CMB of $\ell>2\times 10^6$ which is many orders of magnitude greater than the observable CMB intrinsic anisotropy which due to a combination of Silk damping and 
 foregrounds is unobservable for $\ell>3000$.
 Note that $c_s^2$ is not the square of the photon/baryon fluid sound speed which is of order a $1/3$ during tight coupling, where we are using units with the speed of light set to unity.  Before last scattering,
 the divergence of the photon fluid velocity is sourced by the divergence of the baryon velocity. The photon's density and velocity perturbations are not sourced by the baryon density perturbations. Hence, except on negligibly small scales, the compensated isocurvature mode has all perturbations remaining zero except for $\delta \rho_b$ and $\delta \rho_{\CDM}$ which are initially non-zero but do not grow as over all there is no perturbation to the metric. It follows that the observed CMB fluctuations are unaffected by the presence of a curvaton generated compensated isocurvature mode.

\section{Effect of Compensated Isocurvature on Galaxy Surveys}
\label{effectsgal}
\begin{figure}
\includegraphics[width=7.7cm]{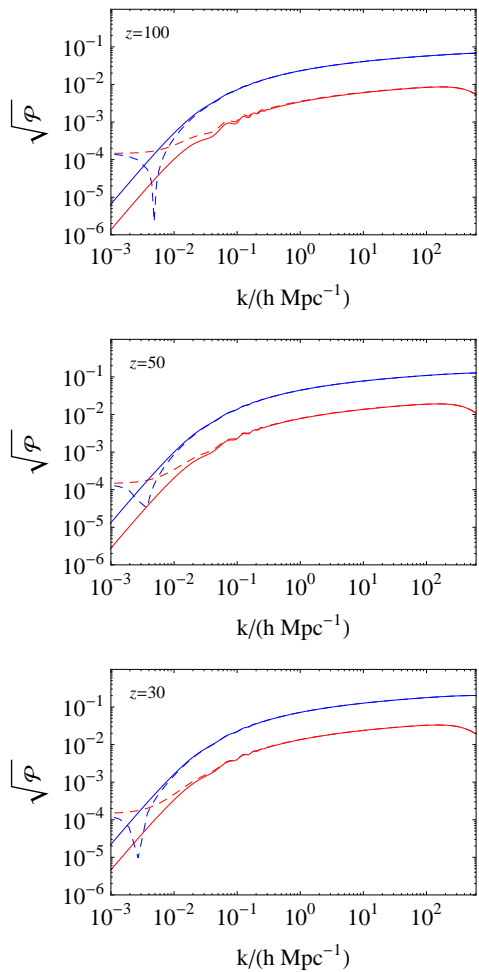}
\caption{\label{spectrum}  The evolution of the dimensionless power spectrum of the baryons (red) and the CDM (blue). The adiabatic (solid) and adiabatic + curvaton compensated isocurvature mode (dashed) are also plotted. The baryon and CDM power spectrum have been scaled by $\Omega_b / \Omega_m$ and $\Omega_{\CDM} / \Omega_m$ respectively, where $\Omega_m=\Omega_{\CDM}+\Omega_b$.  }
\end{figure}

The dimensionless power spectra is given by
\beq
\P\equiv (k^3 / 2\pi^2) \left<\delta \rho(k)^2 \right>/\rho^2\, .
\eeq
It is plotted for the baryon and CDM perturbations  in Fig.~\ref{spectrum} for scales smaller than the Hubble horizon.  
The power spectra for this and the other figures in this article were evaluated using the {\em CAMB sources\/}\footnote{http://camb.info/sources/} program which solves the Boltzmann equation for the photon distribution function numerically \cite{lewcha07}. 
Our fiducial adiabatic model has WMAP5 maximum likelihood parameters \cite{dunkley08}: 
$
\Omega_b h^2=0.0227,\Omega_{\CDM}h^2=0.108,
n=0.961,\tau=0.089, 
\left. {\cal P}_\zeta \right|_{k={1\over 500}\himpc }=2.41\times10^{-9}, h=0.724$ where $\left. H\right|_{z=0}=100h\,$km$\cdot$s$^{-1}\cdot$Mpc$^{-1}$,
 $n$ is the spectral index, and $\tau$ is the optical depth. The compensated mode for the curvaton model is taken to be $\I=-3$ which corresponds to the case of the CDM being created before curvaton decay and baryon number being created by curvaton decay. The compensated mode where baryon number is created before curvaton decay and CDM by curvaton decay corresponds to $\I=3 \Omega_b /\Omega_c=0.63$ and is harder to detect.

 The compensated mode density perturbations are determined by the isocurvature perturbation, \eqs{comp1}{comp}, which in the curvaton model has the same primordial spectral index as the curvature perturbation (${\cal P}\propto k^{n-1}$). The compensated mode density perturbations do not change with time except on scales smaller than the baryon pressure scale corresponding to $k\sim 200\himpc$ at the redshifts probed in Fig.~\ref{spectrum}.
For scales larger than the Hubble horizon at  radiation/matter equality ($k< 0.01h$Mpc$^{-1}$) the adiabatic mode density perturbations are determined by the curvature perturbation and from Poisson's equation  have the form
 ${\cal P}\propto k^{n+3}/(1+z)^2$. For larger $k$, there is a reduction in the slope due to the radiation pressure suppressed growth of sub-horizon scales  during the radiation era. During the matter era all scales grow as $(1+z)^{-1}$ in the adiabatic mode density perturbation.
It follows that the contribution of the compensated mode to the square root of the normalized power spectrum decreases with scale and redshift as $k^{-2}(1+z)$ for  $k<0.01h$Mpc$^{-1}$ and similarly but with slightly less negative slope in $k$ for $k>0.01h$Mpc$^{-1}$.
 The total density perturbation is the sum of the compensated mode and the adiabatic mode density perturbations. The CDM compensated mode density perturbation is completely negatively correlated with the adiabatic mode CDM density perturbation on large scales.
The downward spike seen in the {$\sqrt{\cal P}$} between $k=10^{-3}$ and $10^{-2}\himpc$ is  where the adiabatic and compensated mode CDM density perturbations completely cancel each other.

 The number of gravitationally collapsed halos of mass $M$
is determined by the variance of the non-relativistic matter field (evaluated using linear theory) smoothed with a top hat filter on a scale $R\sim  h^{-1}$Mpc$\left(M/10^{12} M_\odot \right)^{1/3}$ \cite{presch74}.
The number of objects is exponentially suppressed when the corresponding smoothed variance is less than about $1.686^2$ \cite{presch74}. The variance of the smoothed matter field can be approximately read off from the dimensionless power spectrum at scale $k\sim 1/ R$. The first stars form in halos of mass $10^6 M_\odot$. This corresponds to $k \sim 10^{2}h^{-1}$Mpc. On the other end of the scale, clusters of galaxies can have masses up to about $10^{15} M_\odot$ which correspond $k\sim 10 h^{-1}$Mpc.
and tend to form about $z\sim 0$ to  2. As can be seen from Fig.~\ref{spectrum}, the effect of the compensated mode is negligible for $k>0.1 h^{-1}$Mpc for $z\leq100$. It follows that a curvaton generated compensated isocurvature mode will have a negligible effect on collapsed gravitational structures and so is not  detectable by galaxy surveys either.

\section{Effect of Compensated Isocurvature on 21 cm Observations}  
\label{effects21}
\begin{figure}
\includegraphics[width=7.7cm]{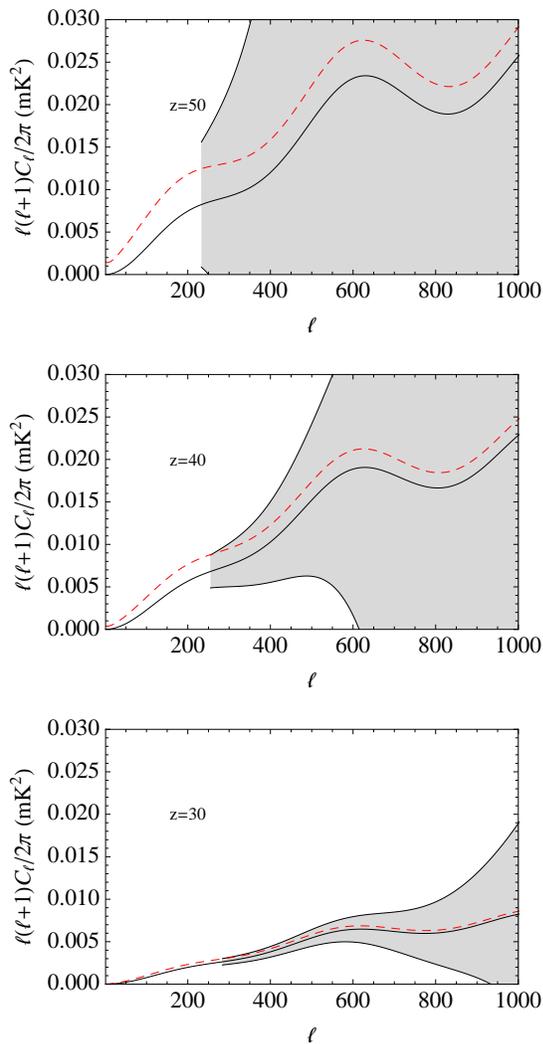}
\caption{\label{errorbars}  
The 21 cm angular power spectrum for the adiabatic mode is plotted (black, solid) with one sigma error bars, using $\Delta \ell =1$.  The signal from the adiabatic plus the curvaton  compensated isocurvature mode is also plotted (dashed, red). The error bars are for a  20~km$^2$ FFT telescope and a band width of 8 MHz.  They start at the minimum value of $\ell$ which is detectable once foregrounds have been taken into account. }
\end{figure}

Absorptions in the observed CMB  at wavelengths 
\beq 
\lambda=21.1(1+z) {\rm cm}
\eeq
 can be used to probe the distribution of neutral hydrogen in the early Universe
(see for example \cite{furohbri06} for a review).
The amount of absorption of the CMB is maximized at $z\sim 70$ where there is the biggest
difference between the CMB temperature and the spin temperature of the neutral hydrogen.
If there is an overdensity of baryons, and hence neutral hydrogen, then there will be more absorption. The amount of neutral hydrogen also affects how the spin temperature evolves with time. There are many other subtle effects, such as distortions due to the peculiar velocities of the neutral hydrogen, but these will be subdominant at the bandwidths, redshifts and length scales we look at here \cite{hirata2007,lewcha07}. To a first approximation, the 21 cm observations can be thought of as mapping the density of neutral hydrogen (or effectively baryons) for redshifts in the range $30\leq z\leq 300$. For lower redshifts, 21 cm can be used for probing reionization (see for example \cite{lidz09}) and as a means of doing galaxy surveys (see for example \cite{abdblaraw09}). Also, for $z<30$, the neutral hydrogen can still be mapped to constrain the matter power spectrum, provided one marginalizes over the reionization model (see for example \cite{priloe08,pripie08,mao08}).
These are the goals of current (e.g.\ MWA \footnote{http://www.MWAtelescope.org/}, LOFAR \footnote{http://www.lofar.org/}) and the next generation (e.g.\ SKA \footnote{http://www.skatelescope.org/}) surveys.
But next$+1$ generation surveys may probe the dark ages for $z>30$. As the neutral hydrogen follows the baryon distribution at high redshifts it will have a signal for a compensated isocurvature mode \cite{barloe05,lewcha07}. In Fig.~\ref{errorbars} the 21 cm signal is plotted at different redshifts for the adiabatic and the adiabatic plus compensated isocurvature mode. The error bars are for those scales measurable after foregrounds have been removed and will be discussed in Sec.~\ref{forecasts}.
For $\ell \delta r/r\gg 1$, (corresponding to $\ell\gg20$ in Fig.~\ref{errorbars}) the angular power spectrum can be approximated by \cite{zalfurher04}
\beq
\ell(\ell+1)C_\ell/2\pi \propto {1\over \ell} \left. {\cal P}_b\right|_{k=\ell/r}
\eeq
where $r$ is the comoving distance to the center of the survey, $\delta r$ is the comoving width of the survey, and ${\cal P}_b$ is the dimensionless power spectrum of the baryons. The extra factor of $1/\ell$ accounts for the smoothing effect of the survey window. For the redshifts in Fig.~\ref{errorbars}, $r\approx 10^4 \himpc$. Fig.~\ref{errorbars} shows that the signal to noise is greatest where $\ell\sim 300$, which corresponds to $k\sim 0.03\himpc$. As can be seen in Fig.~\ref{spectrum}, the baryons in  the compensated mode add non-negligible power at these scales and this can be seen in the corresponding compensated isocurvature plus adiabatic mode curves in Fig.~\ref{errorbars}.

\section{Forecasts}
\label{forecasts}
The FFT telescope \cite{tegzal09} is well suited for measuring high redshift 21 cm. It has antennas arranged in a regular grid allowing an FFT to be used when calculating the correlations between antennae.  This greatly reduces the associated computational cost, which otherwise becomes prohibitive for large arrays with many antennae. Additionally, its large field of view means it can be used in drift mode allowing a quarter of the sky to be surveyed.  As seen in Fig.~8 of \cite{tegzal09}, for the large collecting area needed to survey a redshift of $z=50$, the FFT telescope is the cheapest option.
The noise power spectrum is \cite{tegzal09}
\beq
\label{ClnoiseEq1}
\Clnoise = \Cnoise B_{\ell}^{-2},
\eeq
where the beam function is taken to be Gaussian
\beq
B_{\ell}^{-2}={\rm e}^{\theta^2\ell^2}
\eeq
with the resolution given by \cite{tegzal09}
\beq
\theta=\lambda/\sqrt{A}
\eeq
where $A$ is the FFT telescope area.
Also, \cite{tegzal09}
\beq
\label{CnoiseEq2}
\Cnoise 
=\frac{4\pi}{\eta}\frac{\lambda^3\fsky\Tsys^2}{\fcover A\Omega c\tau}.
\eeq
Here $\fsky\equiv\Omegamap/4\pi$ is the fraction of the sky covered by the map, $\Omega$ is the field of view, we have 
introduced the dimensionless parameter $\eta\equiv\Delta\nu/\nu=\Delta\nu\,c/\lambda$ to denote the relative frequency bandwidth, and $c$ is the speed of light.
In {\em CAMB sources\/} the averaging over frequency is done with a Gaussian of standard deviation $\Delta\nu/(2\sqrt{\pi})$ \cite{rohwil04} specified by  ``redshift\_sigma\_Mhz'' in the {\em CAMB sources\/} initialization file. The observation time is denoted by $\tau$, the system temperature by $\Tsys$, and $\fcover$ is the fraction of the area covered by the array antennas.

The one sigma error bars for the angular power spectrum are \cite{tegzal09}
\beq
\label{dClEq}
\Delta C_\l \approx \sqrt{\frac{2}{(2\l+1)\Delta\l\fsky}}\left(C_\l+\Clnoise\right).
\eeq
Removing foregrounds at these high redshifts will be extraordinarily challenging but, if the foregrounds are sufficiently smooth in frequency, it may well be possible.  We will assume that foregrounds can only be removed for $\ell\ge\ell_{\rm min}$ where $\ell_{\rm min}=k_{\rm min} r$ with the minimum wave number corresponding to the bandwidth and given by \cite{wang06}
\beq
k_{\rm min}=2\pi/\Delta r\, .
\eeq
The Fisher information matrix is given by \cite{tegtayhea96}
\beq
\label{fisher}
F_{ij}=\sum_{\ell\ge \ell_{\rm min}} F_{ij,\ell}
\eeq
where 
\beq
 F_{ij,\ell}=\frac{1}{\Delta C_\ell^2}\frac{\partial C_\ell}{\partial p_i}\frac{\partial C_\ell}{\partial p_j} 
\eeq
and  $p_i$ is parameter $i$.

\begin{figure}
\includegraphics[width=7.7cm]{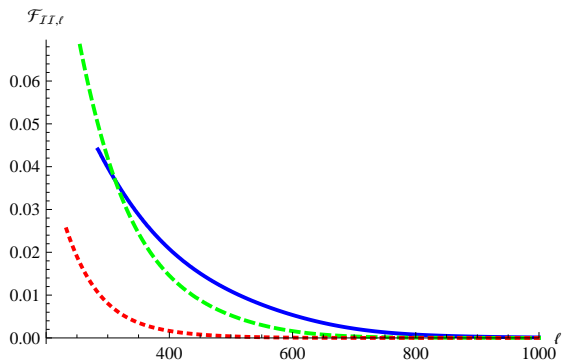}
\caption{\label{fisherfig} The Fisher information of the compensated isocurvature parameter as a function of $\ell$ is plotted for a FFT telescope 
with an area of 20~km$^2$. Redshift 30  
(solid, blue), 40 (dashed green), and 50 (dotted red) are shown.}
\end{figure}

We set the observation time to $\tau=365\times24$~hours and following \cite{tegzel09}, $\Tsys=200 {\rm K} \times [(1+z)/10]^{2.6}$, the field of view to be $\Omega=\Omegamap=\pi$,
and $\fcover=1$. We set the bandwidth to be $\Delta \nu=8$~MHz. The Fisher information for the compensated isocurvature parameter is plotted as a function of $\ell$ in Fig.~\ref{fisherfig}.
As can be seen, at low $\ell$, redshift 40 measurements are more sensitive while at higher $\ell$ redshift 30 measurements are more sensitive. 

\begin{figure}
\includegraphics[width=7.7cm]{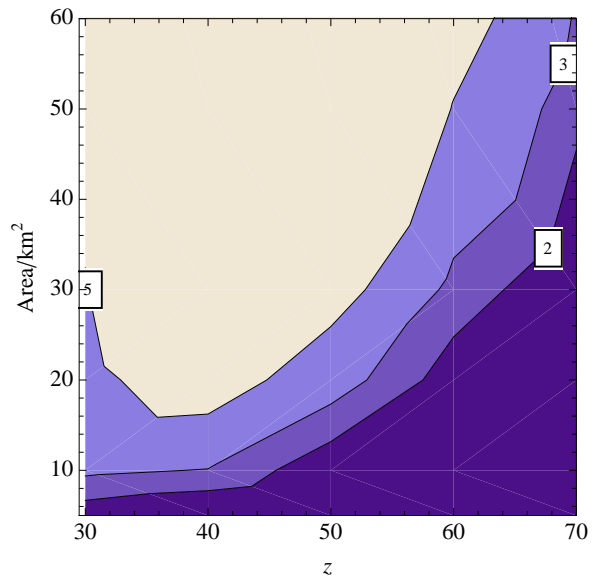}
\caption{\label{cciso} The number of sigma that the curvaton compensated CDM isocurvature mode would be detected is plotted. Each point in the plot is for a 8 Mhz band width experiment.}
\end{figure}

In order to remove degeneracies with the other parameters, we include PLANCK \footnote{http://www.esa.int/SPECIALS/Planck/index.html} CMB forecasts. These do not constrain the compensated isocurvature parameter ($\I$), but they do help to constrain the other parameters which may be degenerate with $\I$. We use the forecasted temperature  and polarization measurements from the 70, 100, 143 and 217 GHz bands of PLANCK, see Sec.~III of \cite{chagorsil08} for more details. The combined Fisher matrix ($F_{21{\rm cm+PLANCK)}}$) is obtained by adding the 21cm and PLANCK Fisher matrices. The forecasted covariance matrix for the parameters is given by inverting the combined Fisher matrix. The number of ``sigma'' for which a curvaton compensated isocurvature mode could be detected is given by $3/(F_{21{\rm cm+PLANCK)}}^{-1})_{\I \I}^{1/2}$ where subscript ${\I \I}$ denotes the element in the row and column corresponding to the compensated isocurvature parameter.
A contour plot of the number of sigma detection, as a function of the area of the FFT telescope and the redshift probed, is plotted in Fig.~\ref{cciso}.
As can be seen, in order to obtain a five sigma detection, a redshift around 40 is optimal and a FFT telescope with an area of at about  20~km$^2$ would be needed. According to \cite{tegzal09}, that area FFT telescope would cost of order several billion US dollars.  However the need to place such an array beyond the Earth's ionosphere, for example on the moon \cite{jester2009}, would raise this cost significantly.

The parameter with the most degeneracy with the compensated isocurvature mode is the baryon density. 
\begin{figure}
\includegraphics[width=7.7cm]{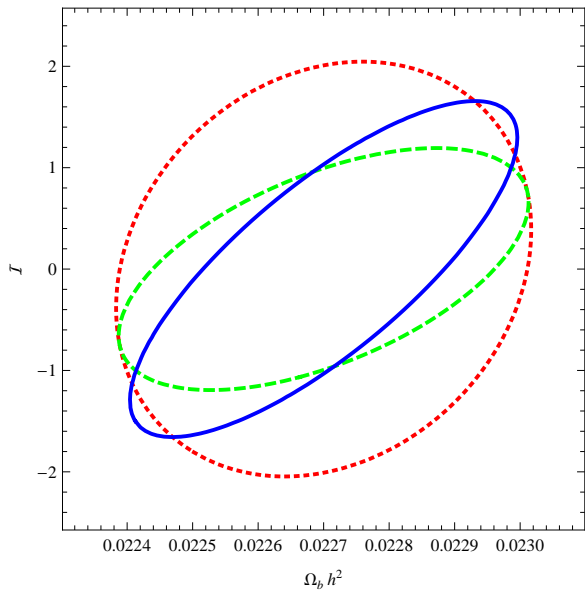}
\caption{\label{IvsB} The contours containing 95\% of the probability distribution for 
a 20~km$^2$ FFT telescope combined with a PLANCK CMB experiment. The axes are for 
the CDM compensated isocurvature mode and the baryon density. All other parameters are marginalized. 
Redshifts 30 (solid blue), 40 (dashed green), and 50 (dotted red) are plotted. The baryon density has the greatest degeneracy with the compensated isocurvature mode.}
\end{figure} 
As can be seen from Fig.~\ref{IvsB}, there is a difference in the slope of the degeneracy for redshifts 30 compared to higher redshifts. This may indicate that combining redshifts will improve the constraints but the results will probably not be significantly improved as the effective area of the telescope can only be optimized for a particular redshift and will then degrade for other redshifts.

\section{Conclusions}
\label{conclusions}
We have shown that it is in principle possible to use 21 cm measurements to detect compensated isocurvature perturbations produced by the curvaton. At redshifts of about 40 the baryon perturbations are sufficiently different from the adiabatic case to give  a detectable 21 cm signal, at the 5 sigma level, provided a FFT telescope of about  20~km$^2$ or larger is used. 

The curvaton model can also produce non-Gaussianity of magnitude $f_{nl}=5/4r$. So for the compensated isocurvature value of
$r\approx 0.17$ a $f_{\rm nl}\approx 7.35$ will be generated. This may be detectable by the PLANCK satellite \cite{komsper01} and potentially even 21 cm observations \cite{cooray06}.

In this article we have looked at a curvaton generated compensated isocurvature mode which from using $\delta \rho_\gamma =0$ in \eq{isodef} and substituting into \eq{comp} gives
 \[
 {\delta \rho_b \over \rho_b} \sim 3 \zeta (\Omega_{\CDM}/\Omega_b)\sim 10^{-3}
 \]
  on large scales.  If compensated modes which have $\delta \rho_b/ \rho_b \sim10^{-1}$
 are considered, then it becomes possible to use many more observational constraints such as
 the scatter in light element abundances due to inhomogeneous big bang nuclear synthesis,
 galaxy cluster gas fraction measurements, B-mode polarization induced from inhomogeneous reionization, and possibly even very large galaxy surveys or high redshift QSO counts \cite{holnoleng09}.
It would be interesting to try and find early Universe models that could induce such a large 
compensated isocurvature mode. 

 In the case of the curvaton generated compensated isocurvature mode,
 there is also likely to be some residual correlated CDM (or equivalently) baryon isocurvature modes unless there is some specific reason for the baryon and CDM isocurvature modes to exactly cancel each other out as in the condition specified in \eqs{comp1}{comp}. 
The WMAP3 data 95\% confidence constraint was \cite{lewis06}  $-0.42 \ge S_{b,\gamma}/\zeta \leq 0.25$. Curvaton model constraints have also been evaluated in WMAP5 \cite{komatsu08,solchahob09}, but they have effectively assumed $S_{b,\gamma} \ge 0$ and so we can't directly use them. If the compensation in \eq{comp} is not exact, there will be a residual 
\beq
{S_{b,\gamma} \over \zeta} = -3 {\Omega_{\CDM} \over \Omega_b} + 3 {1-r \over r}\,.
\eeq
Using the WMAP3 constraints in the above formula we find that $r$ would need to be within about $4\%$ of the compensated value of $r=\Omega_b/\Omega_{\CDM}\approx0.17$. The updated constraints from WMAP5 are likely to be even stronger. So the fine tuning will probably be more severe than $4\%$. However, it is intriguing that the baryon density is so similar to the  CDM  density. This may be a hint that baryogenesis and CDM creation are in some way related and some models have been proposed to account for this (see for example \cite{mcdonald07,kaplan09}).
It would be interesting to investigate if there were some variant of these models that could naturally produce a sufficiently compensated isocurvature perturbation 
to be consistent with current constraints.

\acknowledgments 
 We thank Antony Lewis for helpful discussions.
 CG is supported by the Beecroft Institute for
Particle Astrophysics and Cosmology. JRP is supported by NASA through Hubble Fellowship grant HST-HF-01211.01-A
awarded by the Space Telescope Science Institute, which is operated by the
Association of Universities for Research in Astronomy, Inc., for NASA,
under contract NAS 5-26555. 


\end{document}